\documentclass[preprint,showpacs,preprintnumbers,amsmath,amssymb]{revtex4}

\usepackage{graphicx}
\usepackage{dcolumn}
\usepackage{bm}
\usepackage{hyperref}

\newcommand{\Sr}{Sr$_{14}$Cu$_{24}$O$_{41}$}

\newcommand{\YBaNiO}{Y$_2$BaNiO$_{5}$}
\newcommand{\YBCOx}{YBa$_{2}$Cu$_{3}$O$_{6+x}$}

\begin{document}


\title{Macroscopic quantum coherence of the spin-triplet in the spin-ladder compound \Sr}

\author{J.E. Lorenzo}
 \affiliation{Institut N\'eel-CNRS/UJF, F-38042 Grenoble, Cedex 9, France.}
\author{L.P. Regnault}%
\affiliation{SPSMS, UMR-E 9001, CEA-INAC/UJF-Grenoble 1, MDN, F-38054 Grenoble Cedex 9 , France}%
\author{C. Boullier}%
\author{N. Martin}
\affiliation{SPSMS, UMR-E 9001, CEA-INAC/UJF-Grenoble 1, MDN, F-38054 Grenoble Cedex 9 , France}%
\altaffiliation[Present address ]{BNP-Paribas, Paris, France}
\author{ A.H. Moudden}
\affiliation{Laboratoire L\'eon Brillouin, CEA/CNRS, F-91191 Gif-sur-Yvette Cedex, France}%

\author{V. Saligrama}
\author{C. Marin}
\affiliation{SPSMS, UMR-E 9001, CEA-INAC/UJF-Grenoble 1, IMAPEC, F-38054 Grenoble Cedex 9 , France}%

\author{A. Revcolevschi}
\affiliation{ Laboratoire de Physico-Chimie des Solides, Universit\'e Paris-Sud-11, 91405 Orsay Cedex, France}%

\date{\today}

\begin{abstract}

We report the direct observation by inelastic neutron scattering experiments of a spin-triplet of magnetic excitations in the response associated with the ladders in the composite cuprate \Sr. This appears as a peak at q$_{Q1D}=\pi$ and energy $\Delta_1$ = 32.5 meV and we conjecture that all the triplets making up this conspicuous peak have the same phase and we therefore interpret it as the signature of the occurrence of quantum coherence along the ladder direction between entangled spin pairs. From the comparison with previous neutron and X-ray data, we conclude that the temperature evolution of this mode is driven by the crystallization of holes into a charge density wave in the ladder sublattice. 
\end{abstract}

\maketitle

\textit{Introduction.} \Sr is a compound where the physics of S=1 quasi-one dimensional (Q1D) spin liquids can be best studied \cite{Dagotto1996,Vuletic2006}. It displays a composite crystallographic structure made up of the stacking of two distinct low dimensional Cu-O arrangements. The first sub-system is a 1-dimensional lattice of edge-sharing CuO$_2-$chains and the second one is a 2D sub-system of two-leg Cu$_2$O$_3-$ladders, the stacking direction being the $b-$axis. Lattice parameters for the ladder sublattice are $a$=11.53 \AA, $b$=13.37 \AA, $c_L$ =3.93 \AA. The admixture of both sub-systems originates a superstructure with a nearly commensurate ratio of chain and ladder units along the $c-$direction, 10$c_C \approx$ 7$c_L$ that results in a rather large lattice parameter $c$=27.52 \AA~for the super-cell.

As the uniform magnetic susceptibility vanishes exponentially at low temperatures it has been inferred that the ground state of both subsystems is a non-magnetic spin-singlet \cite{Azuma1994} resulting from the quantum entanglement (antiferromagnetic) of two S=1/2 particles residing at Cu$^{2+}$ positions. Excitations at low temperature are therefore composite bosons of two-particle states. The number of basis states is four: a non-magnetic singlet $\lvert 0 \rangle = \lvert 00 \rangle = \lvert\uparrow\downarrow \rangle  - \lvert\downarrow\uparrow \rangle$  and a magnetic triplet $\lvert 1 \rangle = \left\lbrace  \lvert 10 \rangle, \lvert 11 \rangle, \lvert \bar{1}1 \rangle \right\rbrace =  \left\lbrace \lvert \uparrow\downarrow \rangle + \lvert \downarrow\uparrow \rangle, \lvert \uparrow \uparrow \rangle , \lvert \downarrow \downarrow \rangle \right\rbrace $. Very recently we have verified that these states are exactly the eigenstates of the ground and excited states of the S=1/2 dimers topology \cite{Lorenzo2007}. 
Interestingly \Sr~is a self-doped compound with 6 holes per formula unit ($0.25$ hole per Cu$^{2+}$ ion) and with an unequal distribution of holes between chains and ladders. The total number of holes and its distribution among the two sublattices varies as a function of the substitution Sr by La or Y and by Ca, respectively. Today it is understood that while a charge ordering (CO) appears in the chain sublattice below $\approx $150K, a charge density wave (CDW) is established in the ladder sub-system below $\approx $ 210K in pure \Sr \cite{Vuletic2006, Abbamonte2004}. These characteristic temperatures \cite{Ivek2008} vary from compound to compound and the exact repartition of holes between the two sub-lattices is a matter of very active discussion  \cite{Wohlfeld2007,Rusydi2006}. Therefore this material provides the opportunity to study not only quantum magnetism in low-dimensional compounds but also to investigate how the carriers interact with such a magnetic environment. The answer to this question is relevant in the field of high-T$_C$ cuprates, of which \Sr~is a very close system.

The occurrence of a quantum spin-singlet ground state is revealed in inelastic neutron scattering (INS) experiments through the absence of any elastic or quasi-elastic magnetic scattering and the appearance of a spin gap in the magnetic excitation spectra, at sufficiently low temperatures \cite{Takigawa1998}. In \Sr the chain subsystem exhibits a gap at about 10 meV \cite{Matsuda1999,Regnault1999} whereas the gap amounts to 32.5 meV for the ladder subsystem\cite{Eccleston1998}. 
At low temperatures the spin excitations originating form the chain sub-lattice are concentrated within a narrow band of 1 meV above the gap \cite{Regnault1999,Matsuda1999}, thus implying that the composite bosons are weakly interacting. The spin-dimers of the ladder subsystem are more tightly bounded and strongly interact along the quasi-1D $c-$direction. As shown by inelastic neutron scattering (INS) the dispersion of excitations extends up to 300 meV \cite{Eccleston1998,Matsuda2000,Notbohm2007}. The study of the spin gap associated with ladders below the critical temperature by INS constitutes the goal of this letter and from here on we will disregard the excitations of the chain sub-system.

\begin{figure}
\includegraphics[scale=0.25]{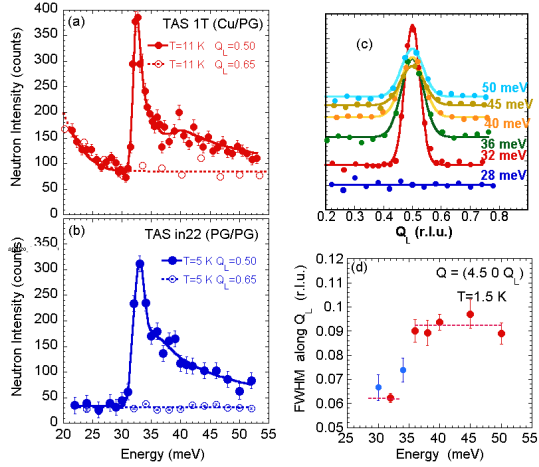}
\caption{\label{fig1} (Color online) (a) Unpolarized neutron energy scan at constant $\bf {Q}$ =(4.5, 0, $Q_L$) showing the gap of the spin fluctuations of the ladder subsystem at T=5K. The scan at $Q_L$=0.65 is taken as a measure of the background and of unresolved nuclear contributions (phonons). These data were taken at TAS 1T with a Cu monochromator. Lines are fits to ad-hoc functions.
(b) Similar scans at the same $Q-$positions taken on TAS IN22 with a PG monochromator. (c) $Q-$scans along the ladder direction ($Q_L$) at different energy transfers and at T=1.5 K. Continuous line is a fit to a Gaussian function. (d) Energy dependence of the FWHM of the scattering shown in figure (c). Points depicted in blue represent the peak at 32.5 meV convoluted with the tails of the resolution function at the corresponding energies.
}
\end{figure}

\textit{Experiment}. INS experiments were carried out on three-axis spectrometer (TAS) 1T at Laboratoire Leon Brillouin (Saclay, France) and on IN22 at the Institute Laue-Langevin (Grenoble, France). The first spectrometer was operated in an unpolarized mode, with a flat Cu(111) as monochromator and a PG(002) as analyzer with only vertically focusing capabilities. Fig.~\ref{fig1}a shows a typical energy scan with a long tail of unresolved non-magnetic excitations below 30 meV and a rather sharp peak at 32.5 meV with Full-Width-Half-Maximum (FWHM) of 1.85 meV. In general the unpolarized neutron scattering technique does not allow for a simple separation between nuclear and magnetic components. However it turns out that the magnetic scattering from the ladders is centered at $Q_L= \pi /c_L$ (or $Q_L$=0.5 in relative lattice units, r.l.u.) positions and it is resolution limited along this direction. Therefore off-center energy scans, $Q_L$=0.65 r.l.u., are a good measure of the nuclear+background component and serve to separate the magnetic contribution from the scans at $Q_L$=0.5. 
TAS IN22 was run in both polarized (Heusler/Heusler) and unpolarized (PG/PG) neutron scattering (monochromator/analyzer) configurations at fixed final neutron energy 30.5 meV. The energy scan in Fig.~\ref{fig1}b was taken on TAS IN22 in the unpolarized mode and shows that the peak at 32.5 meV is  resolution limited as in Fig.~\ref{fig1}a, with a slightly larger FWHM of 2.15 meV.
The first polarized INS experiments were carried out with the standard Helmholtz coils set-up defining the polarization at the sample position. Subsequently some of the measurements have been repeated with the spherical polarization analysis device CRYOPAD \cite{Brown1993}, with which the neutron polarization is more precisely controlled or analyzed. In our measurements, the neutron polarization $\bf{P}$ was set parallel to the scattering vector $\bf{Q}$, configuration for which the magnetic scattering goes into the spin-flip channel and the inelastic nuclear contribution goes into the non-spin-flip channel, the cross-talk (or polarization leakage) between channels depending on the efficiency of the polarizer, the analyser and the $\pi$-flippers. A standard flipping ratio value was 15, as measured on the (0~0~2) Bragg reflection. Counting times were typically of 15 minutes per point at 32.5 meV. It is important to stress that polarized and unpolarized INS data on the sharp feature at 32.5 meV and on the dispersion of excitations at higher energies are quite consistent. 

\textit{Results}. INS studies of the excitations arising from the ladder sub-lattice are rather challenging. The spin-spin exchange interactions along both the rung and the leg of the ladder are mediated through an oxygen atom in a 180$^\circ$ bonding configuration, Cu$-$O$-$Cu. The corresponding superexchange constants are well known to be fairly strong $J_{leg} \approx$ $J_{rung} \approx$ 110-120 meV \cite{Eccleston1998,Matsuda2000} or more recently $J_{leg} \approx$ 186 meV and $J_{rung} \approx$ 124 meV \cite{Notbohm2007}. As a result, the excitations are characterized by a steep dispersion, and large neutron energy transfers (above 100 meV) are required in order to measure the excitations dispersion along the ladder direction. These excitations are almost dispersionless along perpendicular directions, $a^*$ and $b^*$. 

Fig.~\ref{fig1}c displays several $Q_L$ scans across the inelastic excitations shown in Fig.~\ref{fig1}b. No trace for the presence of a double peak structure of distance $\Delta q \approx \sqrt{(E^{2}-\Delta_{1}^{2})}/\pi^{2}J_{leg}$ arising from the dispersion is found in our data, even at 50 meV. At this energy we should expect  $\Delta q \approx $0.035 r.l.u. for $J_{leg} \approx$ 110 meV and $\Delta q \approx$ 0.020 r.l.u. for $J_{leg} \approx$ 186 meV r.l.u., values which are both much smaller than the Q-resolution width at 50 meV, $\Delta_{q}^{res} \approx 0.07$ r.l.u.. In view of the results present in Fig.~\ref{fig1}d, we conclude that the former value of $J_{leg}$ is closer to reality.

Clearly the width of the scattering at 32.5 meV is smaller than that at slightly higher energies, the scattering being always unresolved by the instrument resolution function (Figure~\ref{fig1}d). This change in the width is rather abrupt between $\Delta_1$= 32.5 meV and the the high energy part of the spectra. We understand this feature as the signature of the occurrence of a sharp peak at $\Delta_1$, resolution-limited both in $Q_L$ and in energy, different from the slowly decreasing continuum-like contribution which is observed at higher energy. The sharp resonance-like mode can be considered as some sort of 1D \textit{Bragg peak} centered at a finite energy, that hereon we name as \textit{coherence peak}. Interestingly, this mode bears some similarity with that observed at the spin-gap energy  in \YBaNiO, prototype material of the quasi-one dimensional S=1 Haldane chain \cite{Xu2007}.

\begin{figure}
\includegraphics[scale=0.8]{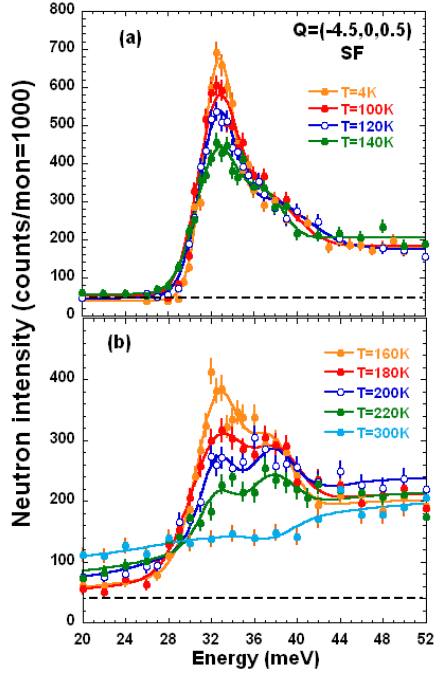}
\caption{\label{fig2} (Color online) (a) Temperature evolution of the energy scans taken at $\bf{Q}$=(-4.5, 0, 0.5) with polarized neutrons between 4 and 140K. Continuous lines are guides to the eye and the broken line is the background level. (b) Temperature evolution of the energy scans between 160K and 300K, with a second peak at $\Delta_2 \approx$ 37 meV.
}
\end{figure}

\begin{figure}
\includegraphics[scale=0.45]{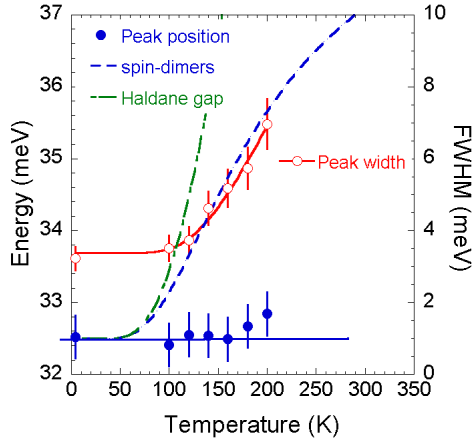}
\caption{\label{fig3} (Color online) Temperature evolution of the coherent peak position and FWHM extracted from the results displayed in Figs. 2. Solid line is the result of the fit of the FWHM by an activated law. Broken (dot-dash) line is the temperature evolution expected for the peak position of the triplet in the spin-dimer model (Haldane-gap).
}
\end{figure}

The temperature dependence of the excitations uncovers the different nature of the magnetic excitations contributing to the gap (coherence peak) and to the dispersion, as shown in Figs.~\ref{fig2}a and~\ref{fig2}b. Similarly to the unpolarized INS results displayed in Figure~\ref{fig1}a, the polarized INS spectra evidence three distinct features. \textit{First}, the intensity of the feature at $\Delta_1$ = 32.5 meV, sharp in energy and in $Q_L$, decreases as the temperature is increased to finally vanish above a characteristic temperature T$_{coh} \approx $ 210K. The peak position remains unchanged within the experimental error bar (Fig.~\ref{fig3}), in striking disagreement with respect to the physics of quantum-dimers  (broken line)\cite{Troyer1994} and Haldane gap (dot-dash line)\cite{Jolicur1994}, and experimental data \cite{Ruegg2005,Zheludev2008}. 
The peak width increases monotonically to reach 7 meV at 200K (Fig.~\ref{fig3}), following an activated behavior ($\propto exp(-\Delta_1/k_BT)$). \textit{Second}, a peak centered at $\Delta_2 \approx$37 meV emerges as the intensity of the 32.5 meV mode decreases, as evident in the scans at temperatures higher than 160 K (see Fig.~\ref{fig2}b). The intensity changes little on increasing temperature, whereas the peak position slightly renormalizes towards higher energies and broadens. \textit{Third}, the long tail of unresolved magnetic excitations is hardly temperature dependent. It corresponds to the dispersion of the triplet of excitations (or triplons), still observable at room temperature since $k_B$T $\ll$ E. Indeed, the thermal behavior of the 32.5 meV peak in \Sr is reminiscent of that of the resonance observed in \YBCOx at optimal doping, namely, a strong decrease of the intensity without hardly any renormalization of the position or broadening.

\begin{figure}[t]
\includegraphics[scale=0.35]{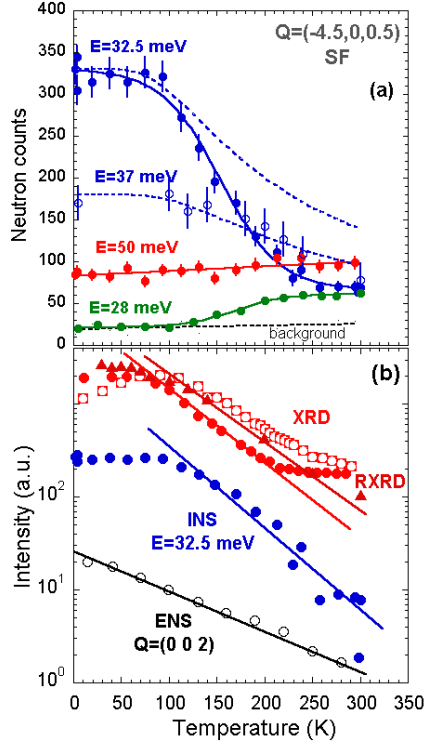}
\caption{\label{fig4} (Color online) (a) Temperature dependence of the intensity at 28, 32.5, 37 and at 50 meV. Solid lines are guides to the eye. Broken lines are the expected behaviors for the singlet-to-triplet population. (b) Linear-log plot of the intensities of the inelastic neutron scattering (INS) experiment with polarized neutrons ; of the elastic neutron scattering (ENS) experiments from reference \cite{Braden2004} ; of the resonant X-ray diffraction experiment (RXRD) \cite{Abbamonte2004}; and of the X-ray diffraction (XRD) \cite{Zimmermann2006}.
}
\end{figure}

Figure \ref{fig4}a shows the temperature dependence of the peak intensity at four characteristic positions in energy: \textit{(i)} 28 meV, just below the coherence peak, that monitors the onset of unresolved energy excitations at the approach of T$_{coh}$, \textit{(ii)} $\Delta_1=$ 32.5 meV, the coherence peak, \textit{(iii)} $\Delta_2=$ 37 meV energy at which an extra peak appears as evident in the intermediate temperature range (Fig. \ref{fig2}b) and \textit{(iv)} at 50 meV. The temperature dependence of the intensity of the coherence peak at 32.5 meV does not display the typical critical behavior of the intensity of an order parameter below a phase transition, (T$_C$-T)$^{2\beta}$, nor the temperature dependence expected for a collection of dimers, $ S(Q,E,T) \approx (exp(\Delta_1/k_BT)-1)/(3+exp(\Delta_1/k_BT)) S(Q,E,T=0)$ (broken line in Fig. \ref{fig4}a). The temperature evolution of the intensity of the coherence peak can be roughly described by an exponential decay I $\approx$ exp(-$\gamma$T) with $\gamma \approx $ 0.017(2)K$^{-1}$ that closely matches the temperature dependence exhibited by the intensity of \textit{(i)} the charge ordering peak of the ladder sub-lattice recently identified in resonant X-ray diffraction (RXRD) \cite{Abbamonte2004},  \textit{(ii)} of the purely (0~0~2) chain reflection measured in elastic neutron scattering (ENS) \cite{Braden2004} and  \textit{(iii)} of the Bragg peaks of both the ladder and chain subsystem measured in X-ray diffraction (XRD) \cite{Zimmermann2006}. 

\textit{Conclusions}. We have studied the spin-gap in the ladder excitations spectra of \Sr by polarised inelastic neutron scattering and identified two distinct resonance modes at energies  $\Delta_1=$ 32.5 meV and $\Delta_2=$ 37 meV. The peak at $\Delta_1$ is sharp, resolution limited in $Q_L$ and energy (the so-called \textit{coherence peak}) and the position is temperature independent. Conversely the intensity decrases very rapidly to dissappear at around 210K. These results are at odds with the current models proposed for low dimensional S=1/2 spin-dimers \cite{Troyer1994} and S=1 Haldane gap compounds \cite{Jolicur1994}. By comparing these results with previous resonant \cite{Abbamonte2004} and non-resonant \cite{Zimmermann2006} X-ray diffraction data we interpret it as resulting from the condensation of holes into a CDW in only about half of the ladders \cite{Wohlfeld2007}. This analysis is consistent with both the presence  of one hole every 20 Cu atoms (average hole density $\eta_{h} \approx$ 0.05) and the value of the CDW propagation vector, $k_{c} \approx 0.2$. The value of $\eta_{h}$ estimated from our INS experiments is about three times smaller than that deduced from the resonant X-ray  diffraction experiments \cite{Abbamonte2004}. At least, our determination of $\eta_{h}$ for the ladder sub-system appears consistent with the almost resolution-limited FWHM in $Q_L$ of the inelastic response at energy $\Delta_1$ and temperatures larger than $T \approx T_{coh}$, which implies $\eta_{h} \sim \pi \Delta_{q}^{res}/3 \approx 0.06$ ($\Delta_{q}^{res}$ in r.l.u.). The scaling feature of the intensities underlines the occurrence of a substantial electronic coupling (Coulomb repulsion) between the ladder and the chain sub-lattices, despite the vanishing small spin and electronic exchange between them. This point has been previously raised in order to explain the RXRD data \cite{Abbamonte2004} and our data agree with this statement. The remaining half of the ladders (weakly hole doped) do not support CDW and thus remain unaffected. 
As the signal at $\Delta_2=$ 37 meV also disappears above $T \approx T_{coh}$ we conclude that this second spin-gap is related to the CDW hole ordering as well. The actual spin-triplet gap of the spin-ladder arrangement may lay at higher energies. The occurrence of long range quantum coherence in an antiferromagnetic quantum gap system has been already reported in the Ni$^{2+}$ spin-1 chain compound \YBaNiO \cite{Xu2007}. Our findings for the resonance at $\Delta_1$ are qualitatively identical to those there reported. Finally we want to emphasize the very close similarities between \Sr and the high-T$_C$ Cu superconductors both in terms of structure and in terms of the signature of the resonance mode coupled to a CDW in the former or affecting the superconducting condensate in the latter.

\textit{Acknowledgment: }Part of this work has been supported by the French ANR project NEMSICOM

\bibliography{Sr14Cu24O41_PRL} 

\end{document}